\documentclass[pre,aps,twocolumn,superscriptaddress]{revtex4-1}

\usepackage{graphicx}
\usepackage{amssymb,amsfonts,amsmath}
\usepackage{color}
\usepackage{ulem}
\usepackage[hidelinks]{hyperref}
\usepackage{bbm}
\usepackage{bbold}
\usepackage{soul}

\newcommand{\rv}{{\mathbf r}}

\newcommand{\Tr}{{\rm Tr}\,}

\newcommand{\pv}{{\bf p}}

\newcommand{\Fv}{{\bf F}}

\newcommand{\msphantom}[1]{$\ldots$}

\newcommand{\eps}{{\boldsymbol \epsilon}}
\newcommand{\epsx}{{\epsilon(x)}}
\newcommand{\taub}{{\boldsymbol \tau}}

\newcommand{\bra}[1]{\langle#1|}
\newcommand{\ket}[1]{|#1\rangle}
\newcommand{\epsnull}{\Big|_{\eps=0}}
\newcommand{\dfunc}[2]{\frac{\delta#1}{\delta#2}}

\newcommand{\unity}{{\mathbb 1}}

\begin{document}

\title{Force balance in thermal quantum many-body systems 
  from Noether's theorem}

\author{Sophie Hermann}
\affiliation{Theoretische Physik II, Physikalisches Institut, 
  Universit{\"a}t Bayreuth, D-95447 Bayreuth, Germany}

\author{Matthias Schmidt}
\affiliation{Theoretische Physik II, Physikalisches Institut, 
  Universit{\"a}t Bayreuth, D-95447 Bayreuth, Germany}
\email{Matthias.Schmidt@uni-bayreuth.de}

\date{31 July 2022, revised version: 16 October 2022,
  submitted to 
  \protect \href{https://iopscience.iop.org/journal/1751-8121/page/Claritons-and-the-Asymptotics-of-ideas-the-Physics-of-Michael-Berry}
           {J.\ Phys.\ A: Math.\ Theor.{\it The Physics of Michael Berry}}}

\begin{abstract}
  We address the consequences of invariance properties of the free
  energy of spatially inhomogeneous quantum many-body systems.  We
  consider a specific position-dependent transformation of the system
  that consists of a spatial deformation and a corresponding locally
  resolved change of momenta.  This operator transformation is
  canonical and hence equivalent to a unitary transformation on the
  underlying Hilbert space of the system. As a consequence, the free
  energy is an invariant under the transformation. Noether's theorem
  for invariant variations then allows to derive an exact sum rule,
  which we show to be the locally resolved equilibrium one-body force
  balance. For the special case of homogeneous shifting, the sum rule
  states that the average global external force vanishes in thermal
  equilibrium.
\end{abstract}

\maketitle 

\section{Introduction}

When investigating global equilibrium properties such as the equation
of state for a given many-body Hamiltonian, the strategies in
classical and quantum statistical mechanical treatments differ
markedly from each other. Obtaining the partition sum in the classical
case requires, in principle, to carry out the high-dimensional phase
space integral over the Boltzmann factor of the Hamiltonian
\cite{hansen2013}. The quantum mechanical analog thereof is the trace
over the Boltzmann factor of the Hamiltonian, where the latter is
viewed as an infinite-dimensional matrix expressed in a suitable basis
of e.g.\ energy eigenfunctions \cite{reif2007}. In both cases, quantum
and classical, the leap from the dynamics of the particle-based
many-body description to the thermal average is both powerful and
abstract. As a result physically meaningful quantities, such as the
pressure, chemical and thermal susceptibilities etc.\ become
systematically available, at least in principle, through derivatives
of the free energy, which is readily available from the partition sum.
On a higher level of detail, locally resolved correlation functions
are available as statistical averages and they characterize the
microscopic structure of the system and allow to obtain global
properties via suitable integration.

On the other hand the concept of forces, while being at the very heart
of mechanics, often receives less attention in both statistical and
quantal contexts.  Nevertheless in the realm of quantum many-body
systems several recent publications
\cite{tarantino2021,tchenkoue2019,schmidt2022rmp} addressed in detail
the force balance relationship on the one-body level of correlation
functions. Here the forces are resolved in position and also in time
in the dynamic case. Tarantino and Ullrich \cite{tarantino2021}
reformulated time-dependent Kohn-Sham density functional theory (DFT)
in terms of the second time derivative of the density. In their
approach forces feature prominently. They argue that the causal
structure of their formulation is more transparent than that of the
standard Kohn-Sham formalism of DFT. Rubio and his coworkers
\cite{tchenkoue2019} have addressed the force balance in several
advanced approximations in DFT.  They state that their approach avoids
differentiability and causality issues and having to carry out the
optimized-effective-potential procedure of orbital-dependent
functionals.

Earlier than these advancements, Tokatly had already honed in on the
force balance relationship in the framework of his time-dependent
deformation functional theory \cite{tokatly2007}.  The theory is based
on considering a hydrodynamic Lagrangian view of quantum many-body
dynamics \cite{tokatly2005one,tokatly2005two}.  The force balance
equation plays a role of a gauge condition that fixes the reference
frame \cite{tokatly2005one}. The approach yields formally exact
equations of motion and conservation laws \cite{tokatly2005one} and it
provided the basis for a geometric formulation of time-dependent
density functional theory \cite{tokatly2005two}.  Ullrich and Tokatly
were then able to address important nonadiabatic effects in the
electron dynamics in time-dependent density-functional theory
\cite{ullrich2006}.

Locally resolved force fields play a prominent role in the recent
power functional framework for many-body dynamics
\cite{schmidt2022rmp}. Besides the time-dependent density profile,
this variational approach includes the locally resolved current and
acceleration distributions as its fundamental physical variables. The
theory has been formulated for classical
\cite{schmidt2013pft,schmidt2018md} and quantum
\cite{schmidt2015qpft,bruetting2019viscosity} systems; reference
\cite{schmidt2022rmp} reviews these approaches and gives much
background. The respective variational equation has the clear physical
interpretation of a nonequilibrium force balance relationship and it
allows to categorize flow and structural forces
\cite{schmidt2022rmp}
and acceleration viscous forces
\cite{schmidt2022rmp}, all of which go beyond the adiabatic forces
that are captured by the dynamical classical DFT
\cite{evans1979,evans1992,evans2016,archer2004ddft,marconi1999ddft}. Standard
DFT is recovered as the equilibrium limit of the power functional
theory \cite{schmidt2022rmp}.

Furthermore, in the classical context, forces were recently put to the
fore in methods to obtain statistically averaged quantities, such as
the density profile of a spatially inhomogeneous system, via computer
simulation of the many-body problem. In his recent review
\cite{rotenberg2020}, Rotenberg gives a clear account of such
force-sampling techniques; see e.g.\ references
\cite{borgis2013,delasheras2018forceSampling,purohit2019} for original
work. On the theoretical side, classical DFT
\cite{evans1979,evans1992,evans2016} offers access to forces via
building the gradient of the Euler-Lagrange minimization equation
\cite{schmidt2022rmp}. An alternative that applies to pairwise
interparticle interactions is the force integral over the two-body
density correlation function \cite{yvon1935,born1946}. The two-body
density is explicitly available within state-of-the-art classical
density functionals, such as fundamental measure theory, see
e.g.\ Ref.~\cite{tschopp2021}.  Two-body density correlation functions
are also central to the recently developed conditional probability DFT
for quantum systems \cite{mccarty2020,pederson2022}.

In prior work we have applied Noether's theorem of invariant
variations \cite{noether1918,byers1998} to the classical statistical
mechanics of particle-based many-body systems
\cite{hermann2021noether, hermann2022topicalReview,
  hermann2022variance,tschopp2022forceDFT}. Rather than starting with
the invariance properties of an action functional, the approach is
based on considering the symmetry properties of appropriate
statistical functionals, such as the partition sum, in order to derive
exact identities. These ``sum rules'' carry clear physical
interpretation as interrelations between forces when starting with
spatial displacement, and between torques when starting with spatial
rotations. Different types of identities result, depending on whether
the elementary free energy is displaced (leading to external force sum
rules), the excess free energy density functional (internal force
identities) or the power functional (memory identities
\cite{hermann2021noether} that connect time direct correlation
functions \cite{schmidt2022rmp}).

We emphasize that Noether's original work \cite{noether1918} is not
restricted to the action integral of a physical system.  She rather
deals with functionals of a general nature, formulating carefully
necessary (and for our practical purposes very mild) assumptions of
analyticity.  Background from an entirely mathematical perspective can
be found in Ref.~\cite{sardanashvily2016}. Descriptions of the
standard application to the action can be found in many sources,
including Refs.~\cite{kosmann-schwarzbach2018} and on a more popular
level Ref.~\cite{neuenschwander2011}. For the classical case, the
differences between the present use in thermal physics and the
standard deterministic form are discussed in
Ref.~\cite{hermann2022topicalReview}.  Briefly, within our present
setting, we require to identify a functional $F[\eps]$ of a
position-dependent vector field $\eps(\rv)$.  On the one hand
$\eps(\rv)$ parametrizes the functional dependence on further fields
(suppressed in the notation). Noether's theorem applies, when despite
this apparent dependence, on the other hand the functional is
invariant under changes of $\eps(\rv)$. Hence trivially $\delta
F[\eps]/\delta \eps(\rv)=0$.  Quite remarkably this leads to a
nontrivial identity, when taken as a concrete recipe for calculation
of the left-hand side explicitly.

The invariant variational techniques have aided the development of a
force-based approach to classical DFT \cite{tschopp2022forceDFT}. Here
the fundamental starting equation is the locally resolved equilibrium
force balance relationship, which (for pairwise interparticle forces)
is a classical result \cite{hansen2013} that dates back to Yvon
\cite{yvon1935}, and to Born and Green \cite{born1946}. The derivation
of this fundamental equation is performed by considering an
inhomogeneous spatial displacement of the entire system, as described
by a vector field $\eps(\rv)$ in three-dimensional space. Together
with a corresponding change of momenta (described in detail below) the
change of variables constitutes a canonical transformation on
classical phase space, and hence it preserves the phase space volume
element \cite{goldstein2002}. The specific form of the transformation
(in particular it being independent of time) also preserves the
Hamiltonian.  Hence the partition sum itself is unchanged under the
transformation and so is the free energy. (The Hamiltonian, via its
associated Boltzmann factor, and the phase space integral are the only
nontrivial ingredients in the partition sum.)  One is hence faced with
an invariant variational problem, as addressed succinctly by Emmy
Noether in her classical work \cite{noether1918}; see reference
\cite{byers1998} for a historical account.

In the present contribution, we demonstrate that Noether's theorem is
applicable to the equilibrium statistical properties of quantum
many-body systems. We present a quantal shifting transformation of the
position and momentum operators that reduces to the transformation of
reference \cite{tschopp2022forceDFT} in the classical case. Quantum
mechanically, the transformation is canonical \cite{anderson1994},
i.e.\ it preserves the fundamental commutator relation between
position and momentum. Such transformations represent unitary
transformations on the Hilbert space of the considered system. The
partition sum is hence invariant under the transformation, as it is
given as the trace of the Hamiltonian's Boltzmann factor, with both
the trace and the Hamiltonian being invariants, as is the case
classically. The result, to first order in the displacement field, is
the locally resolved equilibrium force balance relationship
\cite{tchenkoue2019,tarantino2021,schmidt2022rmp}.  While one could
expect on general grounds that the Noether line of thought would
indeed apply to quantum systems, the details of the derivation differ
markedly from the classical case, and we spell out the details in the
following.

\section{Thermal Invariance Theory}
\subsection{Quantum canonical transformation}
We consider Hamiltonians of the form
\begin{align}
  H = \sum_i \frac{\pv_i^2}{2m} + u(\rv^N) 
     + \sum_i V_{\rm ext}(\rv_i),
     \label{EQHamiltonian}
\end{align}
where the sums run over all particles $i=1,\ldots,N$, with the total
number of particles $N$. All particles possess identical mass $m$ and
each particle $i$ is characterized by its position ($\rv_i$) and
momentum ($\pv_i$) operator. The interparticle interaction potential
$u(\rv^N)$ depends on all particle positions and we use the compact
notation $\rv^N \equiv \rv_1,\ldots,\rv_N$. The system is under the
influence of an external one-body potential $V_{\rm ext}(\rv)$, where
$\rv$ is a generic position variable.

Position and momentum satisfy the fundamental commutator relations
\begin{align}
  [\rv_i,\pv_j] &=  {\rm i} \hbar \delta_{ij}  \unity,
  \label{EQcommutator}
\end{align}
where $\rm i$ is the imaginary unit, $\delta_{ij}$ is the Kronecker
symbol and $\unity$ indicates the $3\times 3$-unit matrix.  The
commutator of two vectors involves transposition according to
$[\rv_i,\pv_j]=\rv_i\pv_j-\pv_j\rv_i^{\sf T}$, where the
multiplication of two vectors is dyadic and the superscript $\sf T$
denotes the transpose of a $3\times 3$-matrix. Hence in component
notation $[r_i^\alpha,p_j^\gamma]=r_i^\alpha p_j^\gamma-p_j^\gamma
r_i^\alpha$, where Greek superscripts $\alpha,\gamma$ denote Cartesian
components of position and momentum.  Equation ~\eqref{EQcommutator}
then reads as $[r_i^\alpha,p_j^\gamma]={\rm i}\hbar \delta_{ij}
\delta_{\alpha\gamma}$. We work in position representation, such that
the momentum operator of particle $i$ is given by $\pv_i = -{\rm i}
\hbar \nabla_i$, where $\nabla_i$ indicates the derivative with
respect to $\rv_i$.

We consider the following transformation of position and momenta
\begin{align}
  \rv_i &\to \rv_i + \eps(\rv_i),
  \label{EQpositionTransformation}\\
  \pv_i &\to \big\{(\unity + (\nabla_i\eps_i))^{-1}\cdot \pv_i
  + \pv_i \cdot (\unity + (\nabla_i\eps_i)^{\sf T})^{-1}\big\}\big/2,
  \label{EQmomentumTransformation}
\end{align}
where $\eps(\rv)$ is a given real-valued three-dimensional vector
field with $\rv$ indicating position, $\eps_i=\eps(\rv_i)$ is a
shorthand notation, and the superscript $-1$ indicates matrix
inversion.  In Eq.~\eqref{EQmomentumTransformation} the gradient
operator $\nabla_i$ acts only on $\eps_i$, as is indicated by the
surrounding parentheses; hence in position representation each entry
of the $3 \times 3$-matrix $(\nabla_i\eps_i)$, which is obtained as a
dyadic product of the vectors $\nabla_i$ and $\eps_i$ acts only as a
multiplication operator on the wave function.  In our notation the
  dot product of a matrix $\sf A$ and a vector $\bf x$ is understood
  in the standard way as $({\sf A}\cdot{\bf
    x})_\alpha=A_{\alpha\gamma}x_\gamma$, with summation being implied
  over the repeated index $\gamma$. For convenience we also define
  this product with the reversed order of factors as $({\bf
    x}\cdot{\sf A})_\alpha=x_\gamma A_{\gamma\alpha}$. This appears in
  the second term in the sum in Eq.~\eqref{EQmomentumTransformation}.

We assume throughout that the vector field $\eps(\rv)$ is such that a
bijection is established between old and new coordinates. Hence the
transformations \eqref{EQpositionTransformation} and
\eqref{EQmomentumTransformation} need to be invertible. (A poignant
counterexample is $\eps(\rv)=-\rv$, which renders
Eq.~\eqref{EQpositionTransformation} to be $\rv_i\to 0$ and the matrix
inversion in the momentum transformation
\eqref{EQmomentumTransformation} becoming ill-defined.)  The momentum
transformation is the self-adjoint version of the classical phase
space transformation considered in Ref.~\cite{tschopp2022forceDFT},
which is in linear order, as given in Ref.~\cite{tschopp2022forceDFT},
simply $\pv_i \to \pv_i - (\nabla_i \eps_i) \cdot\pv_i$. Here there is
no need to pay attention to the ordering of terms, as the classical
phase space variables commute with each other. The finite version
thereof is $\pv_i \to (\unity + \nabla_i \eps_i)^{-1}\cdot\pv_i$, as
can be shown via a generating function that via differentiation yields
the transformation equations.  Equation
\eqref{EQmomentumTransformation} is obtained as the arithmetic mean of
this expression and its adjoint.  Very briefly, the classical
generator (see the appendix of Ref.~\cite{tschopp2022forceDFT}) is
${\cal G}=\sum_{i=1}^N \tilde\pv_i\cdot(\rv_i+\eps(\rv_i))$, and the
transformation equations are obtained via the identities $\tilde\rv_i
= \partial {\cal G}/\partial \tilde\pv_i$ and $\pv_i = \partial {\cal
  G}/\partial \rv_i$.

We expand the inverse matrix in Eq.~\eqref{EQmomentumTransformation}
to linear order in the gradient of the displacement field according
to: $(\unity+(\nabla_i\eps_i))^{-1}=\unity-(\nabla_i\eps_i)$, where
terms of the order $(\nabla_i\eps_i)^2$ and higher have been omitted.
Using this expansion, Eq.~\eqref{EQmomentumTransformation} in
component notation is: $p_i^\alpha \to p_i^\alpha - \sum_\gamma
\big\{(\nabla_i^\alpha \epsilon_i^\gamma) p_i^\gamma + p_i^\gamma
(\nabla_i^\alpha\epsilon_i^\gamma)\big\}/2$, which when resorting back
to vector notation is:
\begin{align}
  \pv_i \to \pv_i - \big\{(\nabla_i\eps_i)\cdot\pv_i
  + \pv_i\cdot (\nabla_i\eps_i)^{\sf T}\big\}/2.
  \label{EQmomentumTransformationLinear}
\end{align}
We first ascertain that the new coordinates $\tilde\rv_i$ and new
momenta $\tilde\pv_i$, as defined by the right-hand sides of the
transformation \eqref{EQpositionTransformation} and
\eqref{EQmomentumTransformation}, also satisfy canonical commutation
relations. We start with the prominent case of position and momentum:
\begin{align}
  [\tilde\rv_i,\tilde\pv_j] &=
  [\rv_i+\eps_i, \notag\\&\qquad
    \pv_j - \big\{(\nabla_j\eps_j)\cdot\pv_j
    + \pv_j\cdot (\nabla_j\eps_j)^{\sf T}\big\}/2]
    \label{EQcommutatorTransformedResultPreliminary2}\\
  &= [\rv_i,\pv_j]
  +[\eps_i,\pv_j]\notag\\&\qquad
  -[\rv_i,(\nabla_j\eps_j)\cdot\pv_j]/2
  -[\rv_i, \pv_j\cdot (\nabla_j\eps_j)^{\sf T}]/2
  \label{EQcommutatorTransformedResultPreliminary}\\
  &= {\rm i} \hbar \delta_{ij} \unity,
  \label{EQcommutatorTransformedResult}
\end{align}
where we have truncated in
\eqref{EQcommutatorTransformedResultPreliminary} at linear order in
the displacement field and its gradient. As the left-hand side of
\eqref{EQcommutatorTransformedResultPreliminary2} involves no coupling
between different particles, it is straightforward to see that for
distinct particles, $i\neq j$, the result vanishes, as is indeed the
case in Eq.~\eqref{EQcommutatorTransformedResult}. For $i=j$ we use
the explicit form of the momentum operator $\pv_i=-{\rm
  i}\hbar\nabla_i$ to find that the second term in
\eqref{EQcommutatorTransformedResultPreliminary} is
$[\eps_i,\pv_i]={\rm i}\hbar (\nabla_i\eps_i)$. This contribution is
precisely cancelled by the sum of the third and the fourth term in
\eqref{EQcommutatorTransformedResultPreliminary}, which can be shown
to have the form $-[\rv_i,(\nabla_i\eps_i)\cdot\pv_i]
=-(\nabla_i\eps_i) \cdot ({\rm i}\hbar\unity)=-{\rm
  i}\hbar(\nabla_i\eps_i)$. Hence the first term in
\eqref{EQcommutatorTransformedResultPreliminary} alone gives the
result \eqref{EQcommutatorTransformedResult} upon using the
fundamental commutator~\eqref{EQcommutator}.

For completeness, the new variables also satisfy
$[\tilde\rv_i,\tilde\rv_j]=0$ and $[\tilde\pv_i,\tilde\pv_j]=0$. The
former relationship is trivial, as in position representation only
coordinates are involved according to the transformation
\eqref{EQpositionTransformation}. The momentum identity can be worked
out straightforwardly, as we show in appendix
\ref{SECappendixMomentumMomentum}.  Furthermore the new degrees of
freedom are self-adjoint operators. For the positions this is trivial,
as we we have $\tilde\rv_i^\dag=\rv_i^\dag +\eps(\rv_i)^\dag =
\rv_i+\eps(\rv_i)\equiv\tilde\rv_i$, because $\eps(\rv)$ is
real-valued. For the momenta: $\tilde\pv_i^\dag = \pv_i^\dag
-\{(\nabla_i\eps_i)\cdot\pv_i+\pv_i\cdot(\nabla_i\eps_i)^{\sf
  T})\}^\dag/2=\pv_i-\{\pv_i\cdot(\nabla_i\eps_i)^{\sf T}
+(\nabla_i\eps_i)\cdot\pv_i \}/2\equiv \tilde\pv_i$.  For completeness
we demonstrate that the transformation is quantum canonical beyond
linear order in appendix \ref{SECappendixMomentumPositionFinite}.

\subsection{Functional derivatives by local shift}

Having ascertained that the new variables form a sound basis for the
description of the quantum mechanics, we wish to illustrate the effect
of the transformation on the system.  The following considerations
will be an essential ingredient in the thermal physics addressed
further below. We wish to investigate the effect on the Hamiltonian
$H[\eps]$, which is obtained by applying the operator replacements
\eqref{EQpositionTransformation} and \eqref{EQmomentumTransformation}
in the form \eqref{EQHamiltonian} of the original Hamiltonian. We
consider the functional derivative of the transformed Hamiltonian with
respect to the displacement field:
\begin{align}
  \frac{\delta H[\eps]}{\delta \eps(\rv)} &=
  \frac{\delta}{\delta\eps(\rv)}\Big(
  \sum_i\frac{{\tilde\pv_i}^2}{2m}
  +u({\tilde\rv}^{N})
  +\sum_i V_{\rm ext}(\tilde\rv_i)\Big).
  \label{EQHepsDerivative}
\end{align}
To make progress, we first consider the fundamental derivatives of the
new position and new momentum with respect to the displacement
field. These are easily obtained as follows:
\begin{align}
  \frac{\delta \tilde\rv_i}{\delta \eps(\rv)}\epsnull &=
  \delta(\rv-\rv_i)\unity,
  \label{EQpositionDerivative}
  \\
  \frac{\delta \tilde\pv_i}{\delta \eps(\rv)}\epsnull &=
  \nabla \big\{
  \delta(\rv-\rv_i)\pv_i
  +\pv_i\delta(\rv-\rv_i)
  \big\}\big/2,
  \label{EQmomentumDerivative}
\end{align}
where $\delta(\cdot)$ denotes the (three-dimensional) Dirac
distribution and the derivatives are taken at vanishing displacement
field, $\eps(\rv)=0$, as is indicated in the notation on both
left-hand sides. The right-hand side of
Eq.~\eqref{EQpositionDerivative} constitutes the density operator of
particle $i$ times the unit matrix. The right-hand side of
Eq.~\eqref{EQmomentumDerivative} is the spatial gradient of the
momentum density operator of particle $i$.  That both correctly
localized operators appear naturally as functional derivatives is an
initial indication that the considered transformation indeed can be
used as a successful probe for the spatially resolved behaviour of the
system.

For completeness, in index notation Eq.~\eqref{EQmomentumDerivative}
reads as
\begin{align}
  \delta \tilde p_i^\alpha/\delta \epsilon^\gamma
  &=\nabla^\alpha(\delta_i p_i^\gamma +
  p_i^\gamma\delta_i)/2,
  \label{EQmomentumDerivativeIndexNotation}
\end{align}
where we have introduced the shorthand notations
$\delta_i=\delta(\rv-\rv_i)$ and
$\epsilon^\gamma=\epsilon^\gamma(\rv)$ and the derivative is again
evaluated at vanishing displacement field (such that higher than
linear powers in the displacement gradient, as they occur in the {\it
  finite} momentum transformation \eqref{EQmomentumTransformation},
vanish).

In order to obtain the functional derivative \eqref{EQHepsDerivative}
of the Hamiltonian we proceed by first differentiating the kinetic
energy.  We defer the detailed calculations to
appendix~\ref{SECappendixKineticEnergyDerivative}, which contains both
the simpler one-dimensional case, where matrix-vector complexities are
absent (appendix~\ref{SECappendixOneDimension}), as well as the
present three-dimensional case
(appendix~\ref{SECappendixThreeDimensions}). The latter calculation,
carried out in index notation in
appendix~\ref{SECappendixThreeDimensions}, gives the following result:
\begin{align}
  \frac{\delta}{\delta\eps(\rv)} \sum_i \frac{\tilde\pv_i^2}{2m}
  \epsnull &=
  \nabla\cdot\sum_i
  \frac{\pv_i\delta_i\pv_i
    + \pv_i\delta_i\pv_i^{\sf T}}{2m}\notag\\
  &\quad - \frac{\hbar^2}{4m}\nabla \nabla^2 \sum_i \delta(\rv-\rv_i).
  \label{EQkineticEnergyDerivative}
\end{align}
We recall that the transpose (superscript $\sf T$) acts on the entire
$3\times 3$-matrix $\pv_i\delta_i\pv_i$ and that the multiplication of
vector and matrix, as is relevant for the divergence, contracts the
vector index with the first matrix index; we recall our description
thereof after Eq.~\eqref{EQmomentumTransformation}.  Equation
\eqref{EQkineticEnergyDerivative} is also given in index notation in
appendix \ref{SECappendixThreeDimensions}.

The first term on the right-hand side of
Eq.~\eqref{EQkineticEnergyDerivative} is directly analogous to the
classical case \cite{tschopp2022forceDFT}, upon viewing the momentum
operators as phase space variables. The second term on the right-hand
side of Eq.~\eqref{EQkineticEnergyDerivative} is genuinely quantum
mechanical, as it is quadratic in $\hbar$ and hence vanishes in the
classical limit $\hbar\to 0$. This contribution can be rewritten upon
expressing the gradient of the Laplace operator as $\nabla
\nabla^2=\nabla^2 \nabla = \nabla\cdot \nabla\nabla$. Then one can
express the second term in Eq.~\eqref{EQkineticEnergyDerivative} as
$-\nabla\cdot\nabla\nabla\sum_i\delta(\rv-\rv_i)\hbar^2/(4m)$. Here
the Hessian of the density operator,
$\nabla\nabla\sum_i\delta(\rv-\rv_i)$, together with the factor
$\hbar^2/(4m)$ forms the quantal kinetic stress contribution
$\nabla\nabla\sum_i\delta(\rv-\rv_i)\hbar^2/(4m)$.

Together with the first term in Eq.~\eqref{EQkineticEnergyDerivative},
which already is of divergence form, we can define the
position-resolved kinetic stress operator (see
e.g.\ Ref.\cite{schmidt2022rmp}) as
\begin{align}
  \hat\taub(\rv) &=
  -\sum_i \frac{\pv_i\delta_i\pv_i + \pv_i\delta_i\pv_i^{\sf T}}{2m}
  +\frac{\hbar^2}{4m} \nabla\nabla \sum_i\delta_i.
  \label{EQkineticStressOperator}
\end{align}

We have hence adopted the convention to include the wave-like
contribution $\hbar^2\nabla\nabla\hat\rho(\rv)/(4m)$ into the
kinematic stress, where $\hat\rho(\rv)=\sum_i \delta(\rv-\rv_i)$ is
the standard form of the one-body density operator. 
The classical kinetic stress is recovered by letting $\hbar\to 0$,
such that $\hat\taub(\rv)$ reduces to $-\sum_i \delta(\rv-\rv_i)
\pv_i\pv_i/m$, where here $\pv_i$ denotes the classical phase space
variable, which trivially commutes with the spatial delta
distribution, and the transpose becomes irrelevant as for the phase
space variable $\pv_i\pv_i=\pv_i\pv_i^{\sf T}$.

We have so far shown that the considered quantum canonical
transformation the functional derivative of kinetic energy with
respect to the displacement field creates the following fundamental
result:
\begin{align}
  \frac{\delta H_{\rm kin}[\eps]}{\delta \eps(\rv)} \epsnull &=
  -\nabla \cdot \hat\taub(\rv).
  \label{EQkineticEnergyDerivativeCompact}
\end{align}
Here we have split the Hamiltonian \eqref{EQHamiltonian} according to
$H=H_{\rm kin}+H_{\rm pot}$, where the potential energy contains the
interparticle and external contributions, $H_{\rm
  pot}=u(\rv^N)+\sum_iV_{\rm ext}(\rv_i)$. As already laid out above,
the functional dependence on $\eps(\rv)$ that is indicated on the
left-hand side of \eqref{EQkineticEnergyDerivativeCompact} arises from
expressing the original positions and momenta in the Hamiltonian
\eqref{EQHamiltonian} via the transformation
\eqref{EQpositionTransformation} and \eqref{EQmomentumTransformation}.
One could view the result \eqref{EQkineticEnergyDerivativeCompact} as
being unexpectedly simple, despite the technical complexity of the
kinematic stress operator $\hat\taub(\rv)$.  Recall that the kinematic
stress occurs in the Heisenberg equation of motion for the one-body
current density \cite{tokatly2005one,tokatly2005two,schmidt2022rmp}
and that it hence constitutes a meaningful physical object in its own
right. That it is created here from the functional derivative of
kinetic energy with respect to the shift field is a strong indicator
that the thermal Noether invariance against the local shifting
transformation given by Eqs.~\eqref{EQpositionTransformation} and
\eqref{EQmomentumTransformation} carries actual physical significance.

Treating the effects of the local displacement transformation on the
potential energy is comparatively easier than the above kinetic energy
consideration, as here only position coordinates are involved and
hence the commutator structure is trivial. The calculation is very
closely analogous to the classical case
\cite{hermann2022topicalReview}.  We obtain
\begin{align}
  \frac{\delta H_{\rm pot}[\eps]}{\delta\eps(\rv)}\epsnull
  &= \frac{\delta u(\tilde\rv^N)}{\delta \eps(\rv)}\epsnull
  + \sum_i \frac{\delta V_{\rm ext}(\tilde\rv_i)}{\delta \eps(\rv)}
  \epsnull\\
  &= \sum_i (\nabla_i u(\rv_N))\delta_i
  +\sum_i (\nabla_i V_{\rm ext}(\rv_i)) \delta_i
  \label{EQpotentialEnergyDerivativeEarly}\\
  &= - \hat \Fv_{\rm int}(\rv) + \hat\rho(\rv)\nabla V_{\rm ext}(\rv),
  \label{EQpotentialEnergyDerivative}
\end{align}
where we have defined the one-body interparticle force density
operator $\hat\Fv_{\rm int}(\rv)=-\sum_i (\nabla_i
u(\rv^N))\delta_i$. The rewriting that involves the external force
field $-\nabla V_{\rm ext}(\rv)$ in
\eqref{EQpotentialEnergyDerivative} is possible as the derivatives
$\nabla_i$ and $\nabla$, as well as the positions $\rv$ and $\rv_i$,
can be identified with each other due to the presence of the delta
function. The negative external force field $\nabla V_{\rm ext}(\rv)$
can then be taken as a common factor outside of the second sum in
Eq.~\eqref{EQpotentialEnergyDerivativeEarly} and the density operator
remains.

Summing up the kinetic energy derivative
\eqref{EQkineticEnergyDerivativeCompact} and the potential energy
identity \eqref{EQpotentialEnergyDerivative} we obtain
\begin{align}
  -\frac{\delta H[\eps]}{\delta\eps(\rv)}\epsnull &=
  \nabla \cdot \hat\taub(\rv) + \hat\Fv_{\rm int}(\rv)
  -\hat\rho(\rv)\nabla V_{\rm ext}(\rv),
  \label{EQHamiltonianDerivative}
\end{align}
which makes explicit that the Hamiltonian generates, via its negative
functional derivative with respect to the displacement field, the sum
of all one-body force density distributions that act in the system.

That the transformation \eqref{EQpositionTransformation} and
\eqref{EQmomentumTransformation} has an effect on the Hamiltonian
could have been expected from the outset, as the transformation has a
nontrivial spatial structure via its dependence on the vector field
$\eps(\rv)$. Hence Noether's theorem seemingly does not apply, due to
the absence of a direct corresponding invariance. In contrast to this
standard application, here we proceed differently and search for an
invariance that applies in thermal equilibrium. This requires an
average to be an invariant rather than the corresponding operator
itself being an invariant, as we lay out in the following.

\subsection{Force balance from thermal Noether invariance}

We hence turn to a statistical mechanical description which we base on
the free energy in the canonical ensemble, expressed as
\begin{align}
  F &= -k_BT \ln Z,
  \label{EQfreeEnergy}\\
  Z &= \Tr e^{-\beta H}
    \label{EQpartitionSumDefinitionInitial}\\
  &=   \sum_n \bra{n} e^{-\beta H} \ket{n},
    \label{EQpartitionSumDefinition}
\end{align}
where $k_B$ denotes the Boltzmann constant and $T$ is absolute
temperature.  The trace in Hilbert space is denoted by $\Tr$ and it is
made explicit in \eqref{EQpartitionSumDefinition} with $\ket{n}$
denoting the complete set of orthonormal eigenstates of $H$ labelled
by index $n$.  (Possibly degenerate energy eigenstates occur multiple
times in the sum.) In more explicit notation, using position
representation, $\ket{n}=\phi_n(\rv^N)$ such that
$\bra{n}\cdot\ket{n}= \int d\rv^N \phi_n^*(\rv^N) \cdot
\phi_n(\rv^N)$, where the integral is over all position coordinates,
$\int d\rv^N\cdot = \int d\rv_1\int d\rv_2\ldots\int d\rv_N\cdot$ and
the asterisk denotes complex conjugation.  Here and throughout, we
assume that the partition sum \eqref{EQpartitionSumDefinition} and
hence the free energy \eqref{EQfreeEnergy} exists, see Giesbertz and
Ruggenthaler's~\cite{giesbertz2019} account of the divergences that
occur in even simple unbounded systems. In our case, we assume (as we
do classically \cite{hermann2022topicalReview}) that the system is
bounded via the influence of appropriate container walls, as modelled
by a corresponding form of the external potential $V_{\rm
  ext}(\rv)$. We hence adopt a pragmatic stance to the existence of
the free energy \cite{mermin2003}.

We expand the free energy \eqref{EQfreeEnergy} in the transformation
parameter according to:
\begin{align}
  \frac{\delta F[\eps]}{\delta \eps(\rv)}\epsnull &=
  -\frac{k_BT}{Z} \frac{\delta Z[\eps]}{\delta \eps(\rv)}\epsnull
  \label{EQfreeEnergyDerivative1}\\
  &=-\frac{k_BT}{Z}
  \Tr
  \frac{\delta e^{-\beta H[\eps]}}{\delta\eps(\rv)} \epsnull
  \label{EQfreeEnergyDerivative2}  \\&
  =\sum_n \frac{e^{-\beta E_n}}{Z}
  \langle n|\frac{\delta H[\eps]}{\delta\eps(\rv)}\epsnull
  |n\rangle,
  \label{EQfreeEnergyDerivative3}
\end{align}
where in Eq.~\eqref{EQfreeEnergyDerivative1} we have used the
definition \eqref{EQfreeEnergy} of the free energy via the partition
sum $Z[\eps]$ of the transformed system.  In
Eq.~\eqref{EQfreeEnergyDerivative2} we have used the form
\eqref{EQpartitionSumDefinitionInitial} of the partition sum and have
exchanged the order of the functional derivative and building the
trace. Equation \eqref{EQfreeEnergyDerivative3} constitutes a thermal
equilibrium average, where $E_n$ denotes the energy eigenvalue
corresponding to the energy eigenstate $\ket{n}$. We have hence
obtained
\begin{align}
  \frac{\delta F[\eps]}{\delta \eps(\rv)}\epsnull 
  &=
  \Big\langle \frac{\delta H[\eps]}{\delta\eps(\rv)}\epsnull
  \Big\rangle_{\rm eq},
  \label{EQfreeEnergyDerivative4}
\end{align}
where on the right-hand side we have used the notation
$\langle\cdot\rangle_{\rm eq}$ to indicate the average over the
canonical ensemble as it occurs in
Eq.~\eqref{EQfreeEnergyDerivative3}; explicitly this is
$\langle\cdot\rangle_{\rm eq}=\sum_n Z^{-1}e^{-\beta E_n}
\bra{n}\cdot\ket{n}$. The identity \eqref{EQfreeEnergyDerivative4} is
remarkable as it indicates that the local transformation
\eqref{EQpositionTransformation} and \eqref{EQmomentumTransformation}
to lowest order in the displacement field generates a well-defined and
physically meaningful thermal average, that of the functional
derivative of the Hamiltonian. This mathematical structure mirrors
closely that of standard partial derivatives of the free energy with
respect to thermodynamic variables, such as e.g.\ obtaining the
entropy via $S=-\partial F/\partial T$.

Before exploiting the specific form of the right-hand side of
Eq.~\eqref{EQfreeEnergyDerivative4} further, we first proceed with the
general invariance argument.  We expand the free energy of the
transformed system to linear order in the displacement field according
to:
\begin{align}
  F[\eps] & =
  F + \int d\rv 
  \frac{\delta F[\eps]}{\delta\eps(\rv)}\epsnull \cdot \eps(\rv)
  \label{EQFexpansion1}\\
  &= F + \int d\rv
  \Big\langle
  \frac{\delta H[\eps]}{\delta\eps(\rv)}\epsnull
  \Big\rangle_{\rm eq}
  \cdot\eps(\rv),
  \label{EQFexpansion2}
\end{align}
where Eq.~\eqref{EQFexpansion1} is the functional Taylor expansion to
linear order and the form \eqref{EQFexpansion2} follows from using
Eq.~\eqref{EQfreeEnergyDerivative4}.

On the other hand, the free energy is an invariant under the quantum
canonical transformation, and hence:
\begin{align}
  F[\eps] &= F,
  \label{EQFinvariant}
\end{align}
where $F$ is the free energy \eqref{EQfreeEnergy} of the original
representation of the system. Equation \eqref{EQFinvariant} holds due
to the fact that canonical transformations are analogous to unitary
transformations on the underlying Hilbert space of the considered
system; see e.g.\ the account given by Anderson \cite{anderson1994}.
We will return to this point below.

From comparison of the Taylor expansion \eqref{EQFexpansion2} with the
free energy invariance \eqref{EQFinvariant} we can conclude that the
linear term in the expansions vanishes identically and it has to do so
irrespective of the form of $\eps(\rv)$. This can only hold provided
that the prefactor vansishes:
\begin{align}
  \Big\langle
  \frac{\delta H[\eps]}{\delta\eps(\rv)}\epsnull
  \Big\rangle_{\rm eq}
  &= 0.
  \label{EQHamiltonianDerivativeVanishes}
\end{align}
Equation \eqref{EQHamiltonianDerivativeVanishes} is a bare consequence
of the invariance of the free energy under the displacement operation,
and it is obtained immediately from $\delta F[\eps]/\delta
\eps(\rv)|_{\eps=0}=0$, as mentioned in the introduction, upon
skipping the Taylor expansion argument expressed in
Eqs.~\eqref{EQFexpansion1} and \eqref{EQFexpansion2}.  The identity
\eqref{EQHamiltonianDerivativeVanishes} holds irrespective of the
precise form of the interparticle interaction potential $u(\rv^N)$ and
of the external potential $V_{\rm ext}(\rv)$ as they appear in the
Hamiltonian \eqref{EQHamiltonian}.

In order to reveal the physical significance of the Noether sum rule
\eqref{EQHamiltonianDerivativeVanishes} we proceed by inserting the
explicit force form of the functional derivative of the Hamiltonian
given by Eq.~\eqref{EQHamiltonianDerivative}, which yields
\begin{align}
  \nabla\cdot\taub(\rv)
  + \Fv_{\rm int}(\rv)
  -\rho(\rv) \nabla V_{\rm ext}(\rv) &=0.
  \label{EQforceDensityBalance}
\end{align}
Here we have introduced the equilibrium averages for the locally
resolved kinetic stress: $\taub(\rv)=\langle \hat\taub(\rv)
\rangle_{\rm eq}$, for the interparticle force density: $\Fv_{\rm
  int}(\rv) = \langle \hat\Fv_{\rm int}(\rv)\rangle_{\rm eq}$, and for
the one-body density distribution: $\rho(\rv)=\langle
\hat\rho(\rv)\rangle_{\rm eq}$.  The force density balance
relationship \eqref{EQforceDensityBalance} is a known exact
equilibrium sum rule, see
e.g.\ Refs.~\cite{tarantino2021,tchenkoue2019,schmidt2022rmp}. Our
derivation demonstrates its origin in the invariance of the free
energy under the quantum canonical transformation
\eqref{EQpositionTransformation} and \eqref{EQmomentumTransformation}.

As a special case we consider a uniform displacement such that
$\eps(\rv)=\eps_0=\rm const$. For classical systems the invariance of
the free energy under such homogeneous displacement leads to the sum
rule of vanishing global external force in thermal equilibrium
\cite{hermann2021noether,hermann2022topicalReview}.  This result
readily translates to the quantum case as follows.

First we obtain the global identity by starting with the locally
resolved force balance relationship \eqref{EQforceDensityBalance} and
integrating over all positions. Two of the resulting integrals vanish,
$\int d\rv \nabla\cdot \taub(\rv)=0$ and $\int d\rv \Fv_{\rm
  int}(\rv)=0$, where the former identity can be shown via integration
by parts and the latter identity is a consequence of the translational
invariance of the interparticle interaction potential: $\int d\rv
\Fv_{\rm int}(\rv) =\int d\rv \langle \hat\Fv_{\rm int}(\rv)
\rangle_{\rm eq} =\langle \int d\rv\hat\Fv_{\rm int}(\rv) \rangle_{\rm
  eq} = \langle \hat \Fv_{\rm int}^{\rm o} \rangle_{\rm eq}=0$. This
holds due to the global force operator vanishing identically:
$\hat\Fv_{\rm int}^{\rm o}= -\sum_i (\nabla_i u(\rv^N))\equiv 0$,
which can be seen straightforwardly by displacing all positions
arguments in $u(\rv^N)$ and observing that this leaves its value
invariant. Explicitly the invariance is
$u(\rv_1,\ldots,\rv_N)=u(\rv_1+\eps_0,\ldots,\rv_N+\eps_0)$, as the
global shift leaves all distance vectors $\rv_i-\rv_j$ unchanged. The
Taylor expansion of the right-hand side is to first order $
u(\rv_1,\ldots,\rv_N)+\eps_0 \cdot \partial
u(\rv_1+\eps_0,\ldots,\rv_N+\eps_0)/\partial\eps_0|_{\eps_0=0}$. The
linear term can be rewritten as $\eps_0\cdot \sum_i \nabla_i
u(\rv^N)$. As this vanishes for any $\eps_0$, the prefactor vanishes
identically which provides the anticipated vanishing of the global
interparticle force.

This reasoning is analogous to Newton's third law, actio equals
reactio, which holds due to the interparticle forces being
conservative.  The standard derivation does not require (nor identify)
the translational invariance. Typically one addresses the special but
important case of pairwise interparticle interactions, with given pair
potential $\phi(r)$ as a function of interparticle distance~$r$. Then
the global interparticle potential energy is $u(\rv^N)=\sum_{k,l}'
\phi(|\rv_k-\rv_l|)/2$, where the primed sum indicates that the case
$k=l$ has been omitted and the factor $1/2$ corrects for double
counting. The global interparticle force is then the (negative) sum of
all gradients, $\hat\Fv_{\rm int}^{\rm o}=-\sum_i \sum'_{k,l}\nabla_i
\phi(|\rv_k-\rv_l|)/2$. Via re-organizing the nested sums one obtains
$\hat\Fv_{\rm int}^{\rm o}= \sum'_{i,k} [\nabla_i \phi(|\rv_i-\rv_k|)
  -\nabla_i \phi(|\rv_i-\rv_k|)]/2 = 0$, identical to the above result
based on invariance.

The only term that remains of Eq.~\eqref{EQforceDensityBalance} after
carrying out the position integral is the external force contribution,
which reads as:
\begin{align}
  -\int d\rv \rho(\rv) \nabla V_{\rm ext}(\rv) &= 0.
  \label{EQexternalForceVansishes}
\end{align}
Equation \eqref{EQexternalForceVansishes} expresses the vanishing of
the average global external force in thermal equilibrium.

Briefly, in our second route to Eq.~\eqref{EQexternalForceVansishes}
we start from the free energy \eqref{EQfreeEnergy} and directly
perform the transformation for the special case of a homogeneous
displacement~$\eps_0$. In this case the momenta are unchanged, as the
gradient of the (constant) displacement field vanishes
identically. Hence kinetic energy is trivially invariant. As laid out
above, the coordinate change does not affect the interparticle
potential energy, as the difference vectors are unaffected. Hence the
only change in the Hamiltonian occurs in the external contribution and
we obtain, following the same argumentation as in the case of
position-dependent shifting, the result
\begin{align}
  -\Big\langle
  \sum_i \nabla_i V_{\rm ext}(\rv_i)
  \Big\rangle_{\rm eq} &= 0,
  \label{EQexternalForceVansishesAlternative}
\end{align}
which is analogous to the previous form
\eqref{EQexternalForceVansishes} upon prepending $1=\int
d\rv\delta(\rv-\rv_i)$ to
Eq.~\eqref{EQexternalForceVansishesAlternative}, moving the delta
function into the thermal average, and identifying the density profile
$\rho(\rv)=\langle\sum_i\delta(\rv-\rv_i)\rangle_{\rm eq}$.

For completeness and as a final step, we make explicit that the
quantum canonical transformation corresponds indeed to a unitary
transformation on Hilbert space, as is relevant for the invariance
\eqref{EQFinvariant} of the free energy under the transformation.  For
the present transformation, to linear order in $\eps(\rv)$, the
transformed Hamiltonian is obtained via functional Taylor expansion in
the following form:
\begin{align}
  H[\eps] = H
  +\int d\rv \frac{\delta H[\eps]}{\delta\eps(\rv)}\epsnull
  \cdot\eps(\rv),
  \label{EQHamiltonianExpansion}
\end{align}
and we recall the explicit one-body force density form
\eqref{EQHamiltonianDerivative} of the functional derivative of the
Hamiltonian. We treat the second term in
Eq.~\eqref{EQHamiltonianExpansion} as a perturbation to the original
Hamiltonian $H$.  (We recall that the thermal average over the
functional derivative $\delta H[\eps]/\delta\eps(\rv)$ directly leads
to the static force density balance
relationship~\eqref{EQforceDensityBalance}.)  Then the transformed
(perturbed) energy eigenstates $\ket{\tilde n}$ are given by
\begin{align}
  \ket{\tilde n} &= \sum_k U_{nk}\ket{k},\\
  U_{nk} &= \delta_{nk}
  +\frac{1-\delta_{nk}}{E_k-E_n}
  \int d\rv  
  \bra{n}
  \frac{\delta H[\eps]}{\delta\eps(\rv)}\epsnull
  \ket{k}
  \cdot\eps(\rv).
  \label{EQtransformationMatrixDirect}
\end{align}
The form \eqref{EQtransformationMatrixDirect} of the matrix that
performs the change of basis follows from applying time-independent
first order perturbation theory, as is appropriate to capture the
effects to linear order in $\eps(\rv)$ that we consider. (We imply
that the prefactor of the integral in
Eq.~\eqref{EQtransformationMatrixDirect} vanishes for $k=n$.)  The
matrix elements of the Hermitian conjugate to
Eq.~\eqref{EQtransformationMatrixDirect} can be obtained via
exchanging indices $n$ and $k$ as
\begin{align}
  U_{nk}^\dag &= \delta_{kn}
  +\frac{1-\delta_{kn}}{E_n-E_k}
  \int d\rv
  \bra{k} \frac{\delta H[\eps]}{\delta\eps(\rv)}\epsnull^\dag\ket{n}
  \cdot\eps(\rv)\\
  &=
  \delta_{nk} - \frac{1-\delta_{nk}}{E_k-E_n}
  \int d\rv
  \bra{n} \dfunc{H[\eps]}{\eps(\rv)}\epsnull \ket{k} \cdot\eps(\rv),
  \label{EQtransformationMatrixInverse}
\end{align}
where to obtain the matrix elements
\eqref{EQtransformationMatrixInverse} we have exploited that the
functional derivative of the Hamiltonian is self-adjoint. We observe
that the sole difference between
Eqs.~\eqref{EQtransformationMatrixDirect} and
\eqref{EQtransformationMatrixInverse} is the minus sign.  Hence we can
see explicitly that unitarity holds, $\sum_k U^\dag_{nk}
U_{km}=\delta_{nm}$ to linear order in $\eps(\rv)$, as was expected on
general grounds~\cite{anderson1994}.

\section{Outlook and Conclusions}

In conclusion we have investigated the consequences of a specific
local displacement operation for the free energy of a quantum
mechanical many-body system. The transformation consists of
position-dependent shifting, as parameterized by a real-valued
displacement (or ``shift'') field, and a corresponding transformation
of the quantum mechanical momentum operator of each particle. The
entirety of the transformation can be viewed as the self-adjoint
version of the corresponding local shifting transformation of the
classical phase space variables \cite{tschopp2022forceDFT}. We have
explicitly shown that the new position and momentum operators are
self-adjoint and that they satisfy the fundamental commutator
relations and hence form a valid and complete set of degrees of
freedom of the considered system. The transformation can be viewed as
a basis change of the underlying Hilbert space of the quantal system
and we have spelled out explicitly the corresponding unitary
transformation between the original and the new basis.

The resulting invariance of the free energy under changes in the
displacement field then leads, following Noether's theorem for
invariant variations, to an exact local identity (``sum rule'') which
we have shown to be the thermal equilibrium force balance. The present
derivation of this known and fundamental result from Noether's theorem
sheds new light on the very nature of the identity.  Existing
derivations are based e.g.\ on the second time derivative of the
one-body density profile \cite{tarantino2021} or, equivalently, on the
first time derivative of the one-body current distribution
\cite{schmidt2022rmp} and then taking the equilibrium limit.

Our results hold for the ground state of the quantum system, as it is
obtained in the limit $T\to 0$ of the free energy of the thermal
system. We have used the canonical ensemble throughout as it captures
the essence of the required thermal physics. We expect the reasoning
to carry over straightforwardly to the grand ensemble with fluctuating
particle number, as the classical canonical
\cite{hermann2022topicalReview} and grand canonical cases lead to
analogous results upon identifying the respective statistical
averages.

Future work could be addressed at investigating how functional
differentiation can be used to obtain quantum sum rules for
higher-body correlation functions, as previously shown for classical
systems \cite{hermann2021noether}. It would be interesting to address
the effects beyond linear order in the displacement field; classically
the variance of the global external force was shown to be constrained
by the external potential energy curvature \cite{hermann2022variance}.
Last but not least it would be worthwhile to find possible
relationships of our displacement field and the strain field that is
central to elasticity theory, see
e.g.\ Refs.~\cite{sprik2021molPhys,sprik2021jcp} for recent work again
in classical systems.

Identifying connections with Tokatly's work
\cite{tokatly2005one,tokatly2005two,ullrich2006,tokatly2007} would be
highly interesting. His approach is more general than what we cover
here, as it allows for the treatment of the dynamical and nonlinear
cases.  Clearly, attempting to generalize our approach to the dynamics
of statistical quantum systems is an exciting and demanding research
task. (We re-iterate that we have here only considered systems in
static thermal equilibrium.)

The Noether argument itself is not restricted to linear
transformations. The second order was shown, for the case of a global
invariance, to relate the variance of fluctuations with the mean
potential energy curvature \cite{hermann2022variance}. Carrying
through this concept for the quantum case is a further very worthwhile
research task.

\appendix
\section{Momentum and position commutators}
\label{SECappendixCommutators}
\subsection{Momentum-momentum commutator}
\label{SECappendixMomentumMomentum}

To derive the commutator of the new momenta, $[\tilde{\textbf{p}}_i,
  \tilde{\textbf{p}}_j]$, we insert the definition of the
transformation \eqref{EQmomentumTransformationLinear} and consider
terms up to linear order in the displacement gradient. In index
notation this reads as follows:
\begin{align}
  2[\tilde{p}^\alpha_i,\tilde{p}^\gamma_j] & = 
  2 [p^\alpha_i,p^\gamma_j] \nonumber \\ 
  &\quad- [p_i^\alpha,(\nabla_j^\gamma \epsilon_j^\delta) p_j^\delta] 
  - [p_i^\alpha, p_j^\delta (\nabla_j^\gamma \epsilon_j^\delta)]
  \nonumber\\
  &\quad - [(\nabla_i^\alpha \epsilon_i^\delta) p_i^\delta, p_j^\gamma]
  - [ p_i^\delta (\nabla_i^\alpha \epsilon_i^\delta), p_j^\gamma].
  \label{eq:NR1}  
\end{align}
The correlator of the original momenta, as it appears in the first
term on the right-hand side, vanishes trivially,
$[p^\alpha_i,p^\gamma_j]=0$.  This identity also allows to take the
operators $p_i^\delta$ and $p_j^\delta$ out of the commutator in the
remaining terms on the right-hand side of Eq.~\eqref{eq:NR1}. We
obtain
\begin{align}
  2[\tilde{p}^\alpha_i,\tilde{p}^\gamma_j] & = 
  - [p_i^\alpha,(\nabla_j^\gamma \epsilon_j^\delta)]  p_j^\delta
  - p_j^\delta [p_i^\alpha,  (\nabla_j^\gamma \epsilon_j^\delta)]
  \nonumber \\ 
  &\quad + [p_j^\gamma,(\nabla_i^\alpha \epsilon_i^\delta)] p_i^\delta
  + p_i^\delta [p_j^\gamma, (\nabla_i^\alpha \epsilon_i^\delta)],
  \label{eq:NR2} 
\end{align}
where we have exploited the anti-symmetry of the commutator,
$[A,B]=-[B,A]$, for rewriting the third and the fourth term on the
right-hand side of Eq.~\eqref{eq:NR2}.

Writing out explicitly the commutator in the first contribution in
Eq.~\eqref{eq:NR2} yields $[p_i^\alpha,(\nabla_j^\gamma
  \epsilon_j^\delta)] = p_i^\alpha (\nabla_j^\gamma \epsilon_j^\delta)
- (\nabla_j^\gamma \epsilon_j^\delta) p_i^\alpha$. Hence the momentum
operator only acts on the gradient of the displacement field,
$(p_i^\alpha \nabla_j^\gamma \epsilon_j^\delta) = - \text{i} \hbar
(\nabla_i^\alpha \nabla_j^\gamma \epsilon_j^\delta)$, where as before
the parentheses indicate that the derivative(s) only act on the
displacement field and we have expressed the momentum operator in
position representation.  Analog manipulation of all remaining
commutators in equation \eqref{eq:NR2} then yields
\begin{align}
  \frac{2\text{i}}{ \hbar}[\tilde{p}^\alpha_i,\tilde{p}^\gamma_j]
  &=
  - (\nabla_i^\alpha \nabla_j^\gamma \epsilon_j^\delta)  p_j^\delta 
  -  p_j^\delta (\nabla_i^\alpha \nabla_j^\gamma \epsilon_j^\delta)
  \nonumber \\
  &\quad + (\nabla_j^\gamma \nabla_i^\alpha \epsilon_i^\delta)
  p_i^\delta + p_i^\delta (\nabla_j^\gamma \nabla_i^\alpha \epsilon_i^\delta). 
  \label{eq:NR3}
\end{align} 
For $i \neq j$ it is now straightforward to see that each term on the
right-hand side of equation \eqref{eq:NR3} vanishes individually: As
the displacement field $\boldsymbol{\epsilon}_i$ only depends on
positions $\textbf{r}_i$, derivatives with respect to $\textbf{r}_j$
vanish for $i \neq j$.  For $i=j$ the first and the third term, as
well as the second and the fourth term, on the right-hand side of
Eq.~\eqref{eq:NR3} cancel each other pairwise, as the derivatives
$\nabla_i^\alpha$ and $\nabla_j^\gamma$ commute.  Collecting the cases
$i=j$ and $i \neq j$ the commutator of the new momentum operators
hence vanishes,
\begin{align}
  [\tilde{p}^\alpha_i,\tilde{p}^\gamma_j] = 0,
  \label{eq:NR4}
\end{align}
which ascertains that the position-dependent momentum transformation
does not generate any spurious terms.

\subsection{Momentum-position commutator for finite transformations}
\label{SECappendixMomentumPositionFinite}

The transformations \eqref{EQpositionTransformation} and
\eqref{EQmomentumTransformation} are canonical not only in linear
order of $\nabla\eps(\rv)$ and $\eps(\rv)$, but also for finite values
thereof.  To demonstrate this property we show that
the canonical commutation relations are satisfied given the finite
transformations \eqref{EQpositionTransformation} and
\eqref{EQmomentumTransformation}.  The position-position commutator
$[\tilde\rv_i,\tilde\rv_j]$ is unchanged compared to the derivation in
linear order. This is due to the transformed position operator
$\tilde\rv_i$ \eqref{EQpositionTransformation} containing no higher
than linear terms in $\eps_i$.

In contrast, the transformed momenta $\tilde\pv_i$ do contain higher
contributions. We express the transformed momentum $\tilde\pv_i$ given
by Eq.~\eqref{EQmomentumTransformation} as an infinite Taylor series
in matrix powers of $(\nabla_i \eps_i)$ as
\begin{align}
  \tilde\pv_i
  =\pv_i +\frac{1}{2}\sum\limits_{n=1}^\infty
  \Big( (-\nabla_i \eps_i)^n \cdot \pv_i 
  + \pv_i \cdot (-\nabla_i \eps_i)^{{\sf T} n}  
  \Big). 
  \label{EQmomentumTranformationFull}
\end{align}
Note that here the order of transposing and raising the power can be
interchanged, i.e.\ $(\nabla_i \eps_i)^{{\sf T} n}=(\nabla_i
\eps_i)^{n{\sf T}}$.

We consider the commutator of position and momentum
$[\tilde\rv_i,\tilde\pv_j]$. Only the case $i=j$ needs to be
considered, since otherwise the commutator vanishes trivially.
Insertion of the transformations \eqref{EQpositionTransformation} and
\eqref{EQmomentumTranformationFull} and exploiting the linearity of
the commutator gives
\begin{align}
  2 [\tilde\rv_i,\tilde\pv_i] &= 2[\rv_i,\pv_i] \label{eq:[r,p]}\\
  &+ \sum\limits_{n=1}^\infty \Big( [\rv_i,(-\nabla_i \eps_i)^n \cdot \pv_i]
  + [\rv_i,\pv_i \cdot (-\nabla_i \eps_i)^{{\sf T} n} ] \Big) \nonumber\\
  &+\sum\limits_{n=0}^\infty \Big( [\eps_i,(-\nabla_i \eps_i)^n \cdot \pv_i]
  + [\eps_i,\pv_i \cdot (-\nabla_i \eps_i)^{{\sf T} n} ] \Big), \nonumber
\end{align}
where the contribution $2[\eps_i,\pv_i]$ is included as the $n=0$ term
of the second series in Eq.~\eqref{eq:[r,p]}.

We rewrite the first term in the first series of Eq.~\eqref{eq:[r,p]}
using index notation:
\begin{align}
  [r_i^\alpha,(-\nabla_i\eps_i)^n_{\beta\gamma} p_i^\gamma] 
  &= (-\nabla_i\eps_i)^n_{\beta\gamma} [r_i^\alpha,p_i^\gamma] 
  \label{eq:nr1a} \\
  &=  (-\nabla_i\eps_i)^n_{\beta\gamma} \text{i} \hbar \delta_{\alpha\gamma}
  \label{eq:nr1b} \\
  &= \text{i} \hbar (-\nabla_i\eps_i)^n_{\beta\alpha}. \label{eq:nr1c}
\end{align}
Here for readability the indices indicating the Cartesian components
of the matrices $(-\nabla_i \eps_i)^n$ are written as subscripts.
The factor $(-\nabla_i\eps_i)^n_{\beta\gamma}$ is local and hence
commutes with position,
$[r_i^\alpha,(-\nabla_i\eps_i)^n_{\beta\gamma}]=0$. Therefore this
term can be taken outside of the commutator in Eq.~\eqref{eq:nr1a}.
We have inserted the usual position momentum commutator in
Eq.~\eqref{eq:nr1b} and evaluated the Kronecker delta in equation
\eqref{eq:nr1c}.

Similarly we express the first term of the second series in equation
\eqref{eq:[r,p]} as
\begin{align}
  [\epsilon_i^\alpha,(-\nabla_i\eps_i)^n_{\beta\gamma} p_i^\gamma] 
  &= (-\nabla_i\eps_i)^n_{\beta\gamma} [\epsilon_i^\alpha,p_i^\gamma]
  \label{eq:nr2a}\\
  &= (-\nabla_i\eps_i)^n_{\beta\gamma} 
  \text{i} \hbar (\nabla_i\eps_i)_{\gamma\alpha} 
  \label{eq:nr2b}\\
  &= - \text{i} \hbar (-\nabla_i\eps_i)^{n+1}_{\beta\alpha}, 
  \label{eq:nr2c}
\end{align}
where again the fact that $\eps_i$ and $\nabla_i\eps_i$ commute allows
to take $(-\nabla_i\eps_i)^n_{\beta\gamma}$ out of the commutator in
equation \eqref{eq:nr2a}.  In Eq.~\eqref{eq:nr2b} we have inserted the
commutator $[\eps_i,\pv_i]=\text{i}\hbar (\nabla_i \eps_i)$.

Recall that both expressions \eqref{eq:nr1c} and \eqref{eq:nr2c} are
part of a sum in equation \eqref{eq:[r,p]}. It becomes apparent that
the $(n+1)$th term of the first sum cancels with the $n$th
contribution of the second sum.  (This amounts to renaming the
summation index in the first sum of equation \eqref{eq:[r,p]} as $n\to
n+1$. Then the first part of both occurring sums become identical up
to a minus sign.)  The only remaining terms are both second
contributions to the sums of equation \eqref{eq:[r,p]}.  These
corresponding transposed terms also cancel each other following an 
analogous argumentation.  Thus no
contribution to the sums in equation \eqref{eq:[r,p]} remains and we
determine the canonical commutator as
\begin{align}
  [\tilde\rv_i,\tilde\pv_j]=[\rv_i,\pv_j] 
  &= \text{i} \hbar \delta_{ij}\unity.
\end{align}
The above considerations generalize
Eqs.~\eqref{EQcommutatorTransformedResultPreliminary2}--%
\eqref{EQcommutatorTransformedResult}
from linear order to the general case.

Show explicitly that the momentum self commutator vanishes,
$[\tilde\pv_i,\tilde\pv_j]=0$ can be done similarly to the treatment
of the position-momentum commutator by explicitly using the
transformation \eqref{EQmomentumTranformationFull}. The corresponding
calculation is straightforward though tedious, and we omit it here.

\section{Local shift derivative of kinetic energy}
\label{SECappendixKineticEnergyDerivative}

\subsection{One dimension}
\label{SECappendixOneDimension}
As a preparation for the general three-dimensional case shown below,
we first consider the simpler case of systems in one spatial
dimension, with position $x_i$ and momentum $p_i=-{\rm i}\hbar
\partial/\partial x_i$ of particle $i=1,\ldots,N$. We consider a
one-dimensional displacement $\epsilon(x)$ of the position coordinate
$x$, such that $x_i\to x_i+\epsilon(x_i)$, in analog to the
three-dimensional case of Eq.~\eqref{EQpositionTransformation}.  The
one-dimensional momentum transformation [corresponding to
  Eq.~\eqref{EQmomentumTransformationLinear}] is $p_i \to p_i -
\{\epsilon'(x_i) p_i+ p_i \epsilon'(x_i)\}/2 \equiv \tilde p_i$, where
the prime denotes the derivative by the argument.

We wish to derive the one-dimensional analogue of
Eq.~\eqref{EQkineticEnergyDerivativeCompact}, which reads as
\begin{align}
  \frac{\delta}{\delta \epsilon(x)}
  \sum_i \frac{\tilde p_i^2}{2m} &=
  \frac{\partial}{\partial x}
  \sum_i \frac{p_i\delta_i p_i}{m}
  - \frac{\hbar^2}{4m} 
  \frac{\partial^3}{\partial x^3} \sum_i \delta_i.
  \label{EQyeah1}
\end{align}
Here the density operator of particle $i$ is defined as
$\delta_i=\delta(x_i-x)$ and the identity holds at $\epsx=0$. The
functional derivative on the left-hand side of Eq.~\eqref{EQyeah1} can
be moved inside of the sum over all particles and we hence need to
consider
\begin{align}
  \frac{\delta \tilde p_i^2}{\delta \epsx}
  &= p_i \frac{\delta\tilde p_i}{\delta \epsx}
  + \frac{\delta \tilde p_i}{\delta \epsx} p_i.
  \label{EQyeah2}
\end{align}
This equality holds to first order in $\epsx$, as we have replaced
$\tilde p_i$ by $p_i$ on the right-hand side.

The first term on the right-hand side of Eq.~\eqref{EQyeah2} becomes
\begin{align}
  p_i \frac{\delta\tilde p_i}{\delta \epsilon(x)}
  & = \partial_x (p_i \delta_i p_i  + p_i p_i \delta_i)/2 \\
  & =  \partial_x\{p_i\delta_i p_i + p_i (p_i\delta_i) + p_i \delta_i p_i\}/2,
\end{align}
where $\partial_x=\partial/\partial x$ is a shortcut notation.  We
have used $\delta p_i/\delta \epsx= \partial_x (\delta_i p_i + p_i
\delta_i)/2$, i.e.\ the one-dimensional analogue of
Eq.~\eqref{EQmomentumDerivativeIndexNotation}, in the first equality
and the product rule of differentiation for $p_i$ in the second
equality. The remaining second term in Eq.~\eqref{EQyeah2} is
\begin{align}
  \frac{\delta\tilde p_i}{\delta\epsilon(x)} p_i
  &=
  \partial_x(\delta_i p_i p_i +p_i \delta_i p_i)/2 \\
  &=
  \partial_x \{p_i \delta_i p_i
  - (p_i\delta_i)p_i + p_i\delta_ip_i\}/2.
  \label{EQcomeOnOn}
\end{align}
The minus sign in Eq.~\eqref{EQcomeOnOn} allows to simplify the sum of
the respective second terms:
\begin{align}
  p_i (p_i \delta_i) - (p_i \delta_i) p_i
  &= 
  (p_i p_i \delta_i)\\
  &=
  -\hbar^2\Big(\frac{\partial^2}{\partial x_i^2}\delta_i\Big)\\
  &=
  -\hbar^2 \partial_x^2 \delta_i,
\end{align}
where in the second equality we have expressed the effect of the
momentum operator on the delta function by the (negative) position
gradient, i.e.\
\begin{align}
  (p_i\delta_i)
  &=
  -{\mathrm i} \hbar \frac{\partial \delta(x-x_i)}{\partial x_i}
  ={\rm i} \hbar \frac{\partial \delta(x-x_i)}{\partial x}.
\end{align}
Collecting all terms yields
\begin{align}
  \frac{\delta \tilde p_i^2}{\delta\epsx}
  &=
  \frac{\partial}{\partial x} 2p_i\delta_i p_i
  -\frac{\hbar^2}{2}
  \frac{\partial^3}{\partial x^3} \delta_i,
  \label{EQyeah4}
\end{align}
and summation over $i$ and division by $2m$ then yields
Eq.~\eqref{EQyeah1}, as desired.

The three-dimensional case covered below is closely related, with the
additional complexity of the matrix and tensor indices interfering
very little with the operator structure.

\subsection{Three dimensions}
\label{SECappendixThreeDimensions}
We wish to derive Eq.~\eqref{EQkineticEnergyDerivative}, which we
reproduce for convenience:
\begin{align}
  \frac{\delta}{\delta\eps(\rv)} \sum_i \frac{\tilde\pv_i^2}{2m}
  \epsnull &=
  \nabla\cdot\sum_i
  \frac{\pv_i\delta_i\pv_i
    + \pv_i\delta_i\pv_i^{\sf T}}{2m}\notag\\
  &\quad - \frac{\hbar^2}{4m}\nabla \nabla^2 \sum_i \delta(\rv-\rv_i).
  \label{EQkineticEnergyDerivativeReproduction}
\end{align}
We use Einstein summation convention over pairs of Greek indices and
after taking the functional derivative set $\eps(\rv)=0$
throughout. We consider the $\gamma$th component of the left-hand side
of Eq.~\eqref{EQkineticEnergyDerivativeReproduction} for particle $i$
only, which yields
\begin{align}
  \frac{\delta }{\delta\epsilon^\gamma}\tilde p_i^\alpha \tilde p_i^\alpha
  &=
  p_i^\alpha \frac{\delta \tilde p_i^\alpha}{\delta\epsilon^\gamma}
  +
  \frac{\delta \tilde p_i^\alpha}{\delta\epsilon^\gamma} p_i^\alpha,
  \label{EQcomeOn}
\end{align} 
where the sum over $\alpha$ (repeated index) generates the square of
momentum, as it occurs in the kinetic energy.  The first term on the
right-hand side, using the explicit form of the transformed momentum
\eqref{EQmomentumTransformationLinear}, becomes
\begin{align}
  p_i^\alpha 
  \frac{\delta\tilde p_i^\alpha}{\delta \epsilon^\gamma}
  & =\nabla^\alpha (p_i^\alpha \delta_i
  p_i^\gamma + p_i^\alpha p_i^\gamma \delta_i)/2 \\
  & =
  \nabla^\alpha\{p_i^\alpha\delta_i p_i^\gamma +p_i^\alpha
  (p_i^\gamma\delta_i) + p_i^\alpha \delta_i p_i^\gamma\}/2,
\end{align}
where we have used Eq.~\eqref{EQmomentumDerivativeIndexNotation} in
the first equality and the product rule of differentiation for the
application of $p_i^\gamma$ in the second equality. The remaining
second term in Eq.~\eqref{EQcomeOn} is 
\begin{align}
  \frac{\delta\tilde p_i^\alpha}{\delta\epsilon^\gamma}
  p_i^\alpha
  &=
  \nabla^\alpha(\delta_i p_i^\gamma p_i^\alpha +p_i^\gamma
  \delta_i p_i^\alpha)/2 \\
  &=
  \nabla^\alpha \{p_i^\gamma\delta_ip_i^\alpha
  - (p_i^\gamma\delta_i)p_i^\alpha + p_i^\gamma\delta_ip_i^\alpha\}/2.
  \label{EQcomeOnOnOn}
\end{align}
The appearance of the minus sign in Eq.~\eqref{EQcomeOnOnOn} allows to
carry out the following cancellation of the respective ``middle''
terms: 
\begin{align}
  p_i^\alpha(p_i^\gamma \delta_i) - (p_i^\gamma \delta_i)
  p_i^\alpha 
  &= 
  (p_i^\alpha
  p_i^\gamma\delta_i)\\
  &=
  -\hbar^2(\nabla_i^\alpha\nabla_i^\gamma\delta_i)\\
  &=
  -\hbar^2 \nabla^\alpha\nabla^\gamma\delta_i,
\end{align}
where in the second step we have rewritten the effect of the momentum
operator on the delta function by the (negative) position gradient,
i.e.\ 
\begin{align}
  (p_i^\gamma\delta_i)
  &=
  -{\mathrm i} \hbar (\nabla_i^\gamma
  \delta_i)={\rm i} \hbar (\nabla^\gamma \delta_i).
\end{align}
Collecting all terms we obtain the overall result for the shift
derivative of kinetic energy,
\begin{align}
  \frac{\delta}{\delta\epsilon^\gamma}
  \sum_i \frac{\tilde p_i^\alpha \tilde p_i^\alpha}{2m}
  \epsnull
  &= \nabla^\alpha\Big\{
  \sum_i \frac{p_i^\alpha\delta_i p_i^\gamma
    +p_i^\gamma\delta_i p_i^\alpha}{2m}\notag\\
  &  \qquad\qquad -  \frac{\hbar^2}{4m} 
  \nabla^\gamma \nabla^\alpha
  \sum_i \delta_i\Big\}
  \label{EQkineticEnergyDerivativeIndexNotation}\\
  & = -\nabla^\alpha \hat\tau^{\alpha\gamma}.
  \label{EQkineticStressDivergenceAppendix}
\end{align}
As desired, Eq.~\eqref{EQkineticEnergyDerivativeIndexNotation} is the
$\gamma$th Cartesian component of
Eq.~\eqref{EQkineticEnergyDerivative} [reproduced above as
  Eq.~\eqref{EQkineticEnergyDerivativeReproduction}] and
Eq.~\eqref{EQkineticStressDivergenceAppendix} is analogous to
Eq.~\eqref{EQkineticEnergyDerivativeCompact}, with tensor contractions
and matrix transpositions expressed in index notation, and the
definition of the kinetic stress operator as given by
Eq.~\eqref{EQkineticStressOperator}.

\acknowledgments

This paper is dedicated to Sir Michael Berry on the occasion of his
80th birthday. MS is grateful for the manifold inspirations and
reliable judgements, in scientific and other matters, that Michael
provided over many years. SH and MS acknowledge useful discussions
with Bob Evans and Daniel de las Heras on the topic of the present
paper. We also thank the Referees for constructive comments and one of
them for pointing out the striking counterexample
$\eps(\rv)=-\rv$. This work is supported by the German Research
Foundation (DFG) via Project No.\ 436306241.

\end{document}